\documentclass[12pt]{article}
\usepackage[dvips]{graphicx}
\usepackage[dvips]{epsfig}

\date{}

\begin{document}

\begin{center}

\textbf{Interplay of magnetic order and Jahn-Teller distortion in a model with strongly correlated electron system}

\vspace{1cm}
     
\textbf{L. Haritha$^{a}$, G. Gangadhar Reddy$^{a,*}$, A. Ramakanth$^{b}$, S.K. Ghatak$^{c}$ and\\ W. Nolting$^{d}$}\\
$^{a}$Department of Physics, Kakatiya University, Warangal - 506 009, India\\
$^{b}$Stanley College of Engineering and Technology for Women, Hyderabad - 500 001, India\\
$^{c}$Department of Physics, IIT Kharagpur, Kharagpur - 721 302, India\\
$^{d}$Institut f{\"u}r Physik, Humboldt-Universit{\"a}t zu Berlin,12489 Berlin, Germany

\end{center}  

\vspace{0.5cm}

\begin{abstract}

The Hubbard model has been employed successfully  to understand  many aspects of correlation driven physical properties, in particular, the magnetic order in itenerant electron systems. In some systems such as Heusler alloys, manganites etc., it is known that, in addition to magnetic order, distortion induced by Jahn-Teller(J-T) effect also exists. In this paper, based on two-fold degenerate Hubbard model, the influence of magnetic order on J-T distortion is investigated. The electron correlation is treated using a spectral density approach and J-T interaction is added to the model. We find that magnetic order and structural distortion coexist at low temperature $T$ for a certain range of electron correlation strength $U$, J-T coupling strength $G$ and band occupation $n$. At $T=0$, for a given $n$ and $U$, magnetic order is present but distortion appears only for a $G$ larger than a critical value. We also studied the temperature dependence of lattice strain and magnetization choosing a $G$ close to the critical value. 

\end {abstract}

\vspace{1.0cm}


\vspace{0.5cm}


\vspace{2.0cm}

\noindent $^*$Corresponding author. Tel.:+919441680795

\noindent E-mail address: gangu\_g@yahoo.com (G. Gangadhar Reddy)

\newpage

\newcommand{\be}{\begin{equation}}
\newcommand{\ee}{\end{equation}}

\section{Introduction}

Strongly correlated electron systems have been at the centre of condensed matter research for quite a long time. One of the oldest models to describe  correlations among itenerant electrons is the Hubbard model. This model is  very well researched using practically all the available theoretical techniques of condensed matter. It is proved to be relevant in many contexts such as itenerant magnetism, metal insulator transitions, high temperature superconductivity etc. On the other hand, the interplay of magnetic and electric phenomena has attracted attention in the context of colossal magneto resistance (CMR), the prototype materials for this being the manganites[1,2]. In these materials it is still not quite clear whether the observed magnetism stems from localized moments ($t_{2g}$-electrons) or from the itenerant ones ($e_g$-electrons). In manganites another essential interaction is the Jahn-Teller (J-T) coupling[3] which causes a structural distortion. The story of manganites is therefore, not complete without including magnetism and J-T coupling. If the magnetism is due to the band electrons and if the band degeneracy can be lifted by J-T coupling, then the study of Hubbard model extended by a J-T term is certainly relevant to these systems. This model is interesting from another point of view also. The emerging field of \textit{orbitronics},  like CMR, also has potential applications. The orbital dynamics of active orbitals is strongly influenced by their interaction with lattice. One way of including electron-lattice interaction is via J-T coupling [4-7]. Here again to be realistic, one has to take the correlation among the orbitals into account, which brings us back to the Hubbard model extended by J-T interaction. Heusler alloys[8,9] are another class of systems where there is a strong interplay between magnetism and lattice distortion. Here of course, the magnetism is that of a local moment system. In this context also, the general effect of magnetic order on lattice distortion is worth studying.

The Hubbard model in general is obviously not exactly solvable. One of the approximations proposed is the spectral density approach (SDA) [10]. In this approach, an ansatz is made that the spectral density is predominantly a two-peak function which is a very reliable ansatz [10,11]. The limitation of this ansatz is that, in this approximation, the quasiparticles are infinitely long living. Earlier investigations have shown that the lifetime of quasiparticles does not qualitatively influence magnetic ordering [12,13]. The J-T coupling can be incorporated into the Hubbard model by renormalizing the band energies. In the present work, the SDA is used to solve the renormalized Hubbard model. The influence of the correlation driven magnetic order on the structural distortion caused by J-T interaction is studied by calculating the magnetization and strain self-consistently.

\section{The model and theory}

The model consists of correlated electrons in a doubly degenerate band interacting with the lattice via the J-T interaction,
\be
H=H_s+H_{JT}+H_L
\ee
where $H_s$ is the  Hubbard Hamiltonian for two degenerate bands and is given by 
\be
H_{s}=\sum_{\alpha,i,j,\sigma}\left(T_{ij}-\mu\delta_{ij}\right)c_{\alpha i\sigma}^{\dagger}c_{\alpha j\sigma} +
      \frac{1}{2} U \sum_{\alpha,i,\sigma}n_{\alpha i\sigma}n_{\alpha i-\sigma} 
\ee
$T_{ij}$ is the hopping integral for hopping of the electrons from lattice site $i$ to $j$. $c_{\alpha i\sigma}^{\dagger}$($c_{\alpha i\sigma}$) is the creation(annihilation) operator for an electron in the $\alpha$-state on the lattice site $i$ with spin $\sigma$.  $\alpha=1,2$ is the band index. $\mu$ is the chemical potential. $n_{\alpha i\sigma} = c_{\alpha i\sigma}^{\dagger}c_{\alpha i\sigma}$ is the number operator. $U$ is the intra-atomic Coulomb repulsion which we take to be the same for both the bands. The intersite hopping integrals $T_{ij}$ are connected to Bloch energies $\epsilon(\mathbf{k})$ by
\be
\epsilon(\mathbf{k})=\frac{1}{N}\sum_{i,j} T_{ij} e^{-i\mathbf{k}
\cdot(\mathbf{R}_i-\mathbf{R}_j)}
\ee
The electron density in the degenerate band couples to the static elastic
strain through the J-T interaction. In the case of a tetragonal distortion, 
this interaction is described by[5-7]
\be
H_{JT}=Ge\sum_{i,\sigma}(n_{1i\sigma}-n_{2i\sigma}) 
=Ge\sum_{\mathbf{k},\sigma}(n_{1\mathbf{k}\sigma}-n_{2\mathbf{k}\sigma}).
\ee
$G$ is the strength of the J-T coupling and $e$ is the lattice strain given by
\be
e=\frac{G}{NC_0}\sum_{i,\sigma}(<n_{1i\sigma}>-<n_{2i\sigma}>)
\ee
where $C_0$ is the elastic constant which we take to be unity for numerical
calculations. It is clear that $H_{JT}$ tries to create a difference 
in the occupation of the
two degenerate bands. The difference in occupation leads to the building up of
the strain. Thus, under suitable conditions, there is a spontaneous splitting of
the bands and building up of the strain which indicates a structural transition. The
building up of the strain, however, leads to an increase in the lattice elastic
energy[4-7] which is given by
\be
H_L=\frac{1}{2}NC_0e^2
\ee
Where $N$ is the total number of atoms. Since this term is a c-number and we are
not looking for the ground state whose energy has to be minimum, we leave this
term out of our consideration. Without resorting to any approximation, we can 
absorb the J-T term into $H_s$ by modifying the band energies for the two bands as
\be
\epsilon_\alpha(\mathbf{k})=\epsilon(\mathbf{k})+(-1)^\alpha G e .
\ee
Where $Ge$, is half the separation between the centres of the
gravity of the split free-bands. With this, the model
Hamiltonian becomes,
\be
H=\sum_{\alpha,\mathbf{k},\sigma}(\epsilon_\alpha(\mathbf{k})-\mu) 
c_{\alpha \mathbf{k}\sigma}^{\dagger}c_{\alpha \mathbf{k}\sigma} +
\frac{1}{2} U \sum_{\alpha,i,\sigma}n_{\alpha i\sigma}n_{\alpha i-\sigma}, 
\ee
which is a two-band Hubbard model. The two bands are coupled through the strain which is introduced J-T by interaction. In order to calculate the strain one has to calculate the single-electron Greens function
\be
G_{\alpha\mathbf{k}\sigma}(E)=\langle\langle c_{\alpha \mathbf{k}\sigma};
c_{\alpha \mathbf{k}\sigma}^{\dagger}\rangle\rangle_E = 
\frac{1}{E-\epsilon_\alpha(\mathbf{k}) +\mu - \Sigma_{\alpha\sigma}(E)}.
\ee
That means one has to calculate the self-energy $\Sigma_{\alpha\sigma}(E)$. In the recent past several approximation schemes have been developed to calculate self-energy. We make use of the self-energy which is obtained[5] by means of the spectral density approach(SDA). As far as magnetic properties are concerned, this approach has proven to be quite reliable.
\be
\Sigma_{\alpha\sigma}(E) = Un_{\alpha -\sigma} + \frac {U^2n_{\alpha -\sigma}(1-n_{\alpha -\sigma})} 
{E + \mu - B_{\alpha - \sigma} - U(1-n_{\alpha -\sigma})}
\ee
The pioneering work of Harris and Lange [11]  to the Hubbard model demonstrates that, in the strong coupling regime, the single-electron spectral density consists of two prominent peaks near the energies $T_0$ and $T_0+U$. This is the starting point for the SDA which uses a linear combination of two weighted $\delta$-functions. Use of $\delta$-functions implies neglect of quasiparticle damping. This is not a serious restriction, at least for strongly coupled systems. It is known that the magnetic stability is adversely affected by finite quasiparticle lifetimes quantitatively but the qualitative behaviour remains the same. The ansatz contains four unknown parameters, namely, the two quasiparticle energies and the two spectral weights, which are fitted by equating the first four, rigorously calculated spectral moments. It turns out that the self-energy given above (Eq.(10)) fulfills the correct high-energy behaviour [12]. The SDA is essentially equivalent to the Roth two-pole approximation for the Greens function [14,15]. This SDA self-energy  can also be obtained by using the Mori-Zwanzig projection technique [16-18].  $B_{\alpha - \sigma}$ is known as the so-called spin-dependent band shift which is decisive for ferromagnetism and is given by
\begin{eqnarray}
B_{\alpha \sigma} &=& T_{\alpha \sigma} - \frac {1} {\pi n_{\alpha \sigma}(1-n_{\alpha \sigma})}
\Im \int_{-\infty}^{+\infty} dE f_-(E)
\Big[\frac{2}{U} \Sigma_{\alpha \sigma}(E) -1\Big] \times {}
\nonumber \\
&& \times   \Big[\Big(E - T_{\alpha \sigma}   +\mu-
  \Sigma_{\alpha \sigma}(E)\Big) G_{\alpha \sigma}(E) -1\Big] {}
\end{eqnarray}
Where $T_{\alpha \sigma}= T_{0} + (-1)^\alpha G e$. $T_{0}$ is the centre of gravity of free Bloch band.  $G_{\alpha \sigma}(E) = \frac{1}{ N}\sum_\mathbf{k} G_{\alpha\mathbf{k}\sigma}(E)$.  From $G_{\alpha\mathbf{k}\sigma}(E)$ one can obtain the density of states by using
\be
\rho_{\alpha\sigma}(E)=-\frac{1}{\pi N}\sum_\mathbf{k} \Im G_{\alpha\mathbf{k}\sigma}(E+i0^+)
\ee
\noindent From the knowledge of the density of states, the expectation
values can be evaluated:
\be
\left<n_{\alpha\sigma}\right>=\int dEf_{-}(E)\rho_{\alpha\sigma}(E)
\ee
Where $f_{-}(E)=1/\left(1+e^{\beta E}\right)$ is the Fermi function with
$\beta=1/kT$.
The chemical potential $\mu$ is fixed by the constraint
\be
n=\sum_{\alpha,\sigma}\left<n_{\alpha\sigma}\right>=constant.
\ee
Now, in terms of the average occupations of the bands
$\left<n_{\alpha\sigma}\right>$, the strain $e$ is given by 
\be
e =\frac{G}{C_0}\sum_{\sigma}\left(\left<n_{1\sigma}\right>-\left<n_{2\sigma}\right>\right)
\ee
The magnetization $m$ is the sum of the magnetizations $m_1$ and $m_2$ of the split bands which are defined by
\be
m_{\alpha} = \left<n_{\alpha \uparrow}\right>-\left<n_{\alpha \downarrow}\right>
\ee
For the numerical evaluation of orbital occupancies and band shift, the $\mathbf{k}$-summation can be conveniently
replaced by an integration over energy using the model density of states:
\be
\rho_{0}(E)=\frac{1}{N}\sum_{\mathbf{k}}\delta(E-\epsilon(\mathbf{k}))
\ee
We solve Eqns.(14), (15) and (11) self-consistently.  In the next section we present
the results.

\section{Results and discussion}

The aim of the present investigation is to study the influence of magnetic order on the structural transition. The model system is characterized by the following parameters: the density of states for free Bloch band which is chosen to be of bcc type [19] with a width  $W=1$, the intra atomic Coulomb repulsion $U$, electron-phonon coupling constant $G$ and the Band filling $n$. For various sets of these model parameters we have evaluated the band occupancies and the strain self-consistently.

First we consider the case of $G=0$ i.e the case of no J-T splitting to find the parameter space in which the system is magnetically ordered. In the absence of J-T distortion, the bands are degenerate and both the bands have same occupations. We determined the $T_c$ which is the temperature at which the spontaneous magnetization becomes zero. In Fig.1 we have plotted $T_C$ as a function of intra-atomic Coulomb repulsion $U$ for various values of band filling $n$. It is noticed that for a given $n$ it requires a minimum U for the onset of magnetic order and the $T_C$ increases with increase of $U$ and gets saturated for a large value of $U$. From the inset of Fig.1 it can be noted that only a certain range of the values of charge carrier concentrations is favourable for magnetism. This can be understood from the $T_C$ dependence of $n$ for $U = 3$ which is plotted in the inset.

\begin{figure}[htb]
\begin{center}
    \epsfig{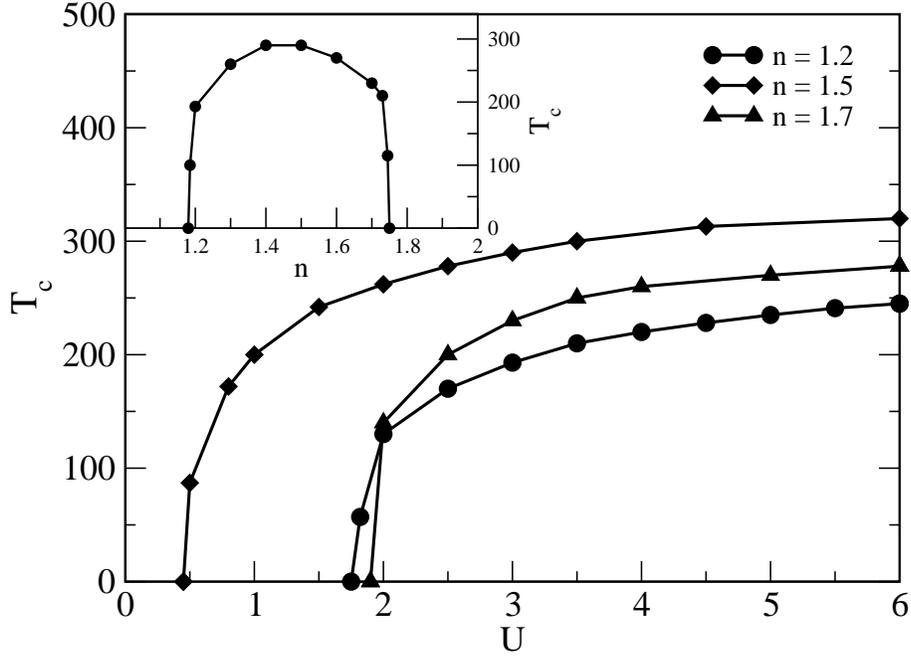}
    \caption{Curie Temperature $T_C$ as a function of Coulomb interaction $U$ for different values of band filling $n$ without Jahn-Teller distortion(i.e. with $G = 0$). The inset shows the $n$ dependence of $T_C$ for $U = 3$.}
\end{center}
\end{figure}

\begin{figure}[htb]
\begin{center}
    \epsfig{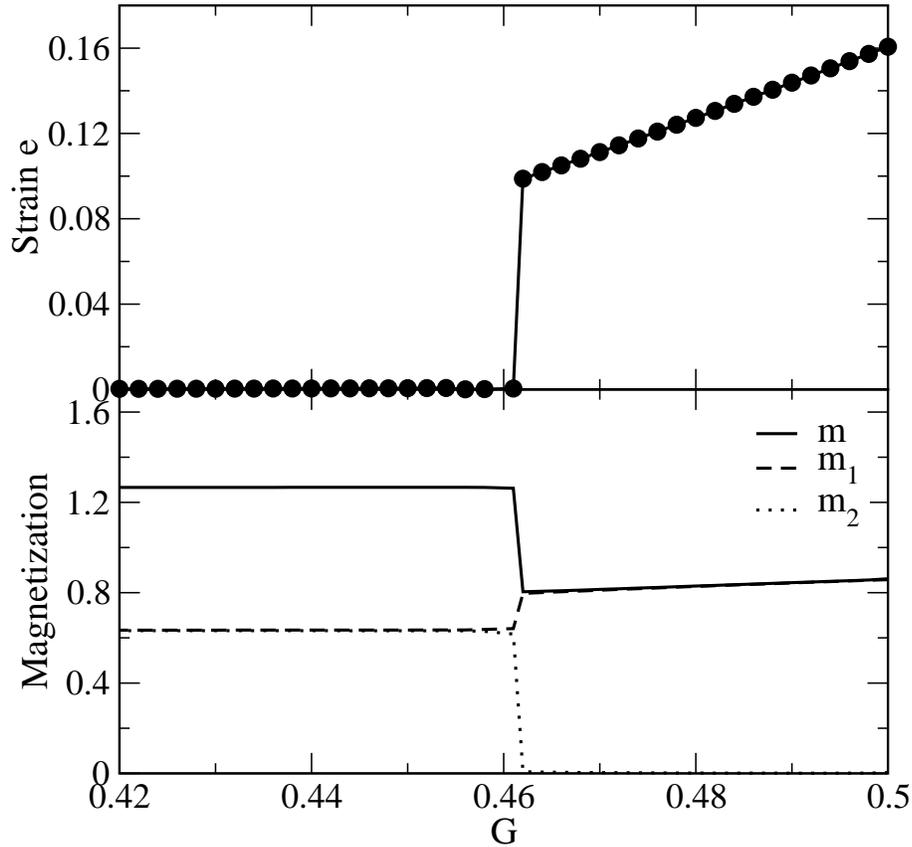}
    \caption{Strain and magnetization as  functions of electron-phonon coupling constant $G$ for band filling $n=1.4$ and $U=3$ at $T=0$.}
\end{center}
\end{figure}

\begin{figure}[htb]
\begin{center}
    \epsfig{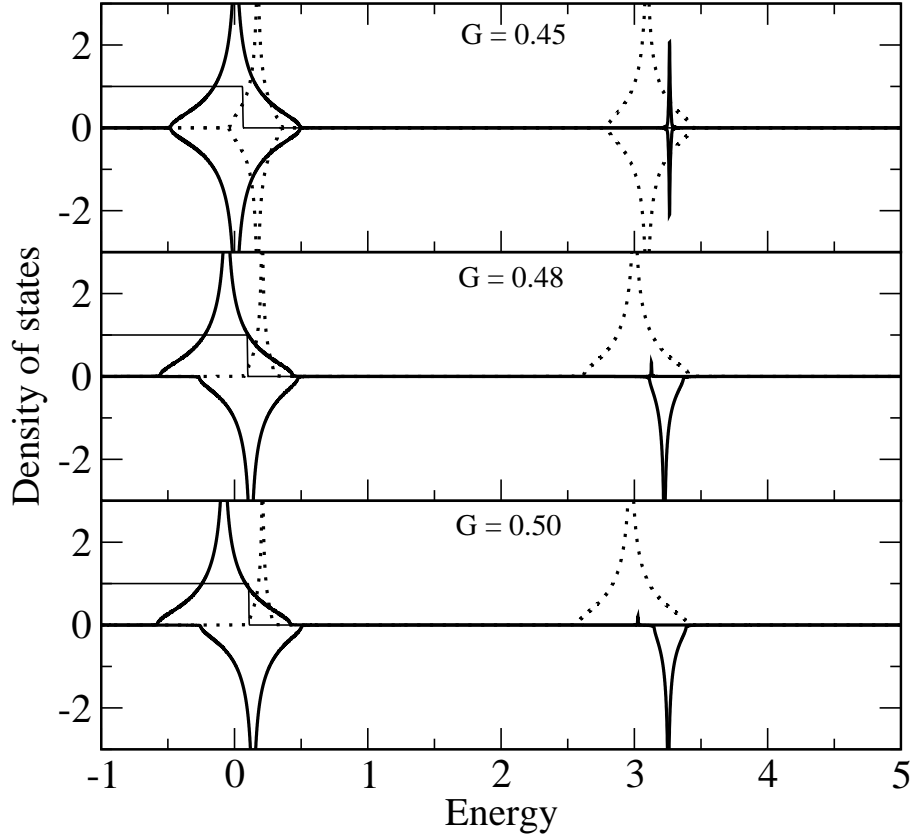}
    \caption{Quasiparticle density of states(for the lower subband in the positive half of the frame and for the upper subband in the negative  half of the frame) as a function of energy for various values of electron-phonon coupling constant $G$. Full lines for spin up and dotted line for spin down bands. Thin line is the Fermi function. $n = 1.4$, $U = 3$ and $T = 0$}
\end{center}
\end{figure}

In order to see the influence of magnetic order on the structural transition we start with the situation where the model system is  magnetically ordered and then switch on the the electron-phonon coupling constant $G$ which makes the system to be eventually distorted. For this we choose the model system with $n=1.4$ and $U = 3$ where the $T_c = 290$ for $G=0$ and calculate $e$ self-consistently from Eq.(15) by progressively increasing the value of $G$ at $T=0$. The result is depicted in Fig.2 (upper part). It is found that J-T splitting does not take place unless the electron-phonon coupling is larger than a critical value $G_c$($\approx 0.462)$. When the value of $G$ is further increased, the strain increases almost linearly. Below the critical value  $G_c$ both the bands are magnetically ordered with equal magnetization($m_1 = m_2 \not= 0$ and $e=0$). However, above the critical value of $G_c$  the magnetization is mostly due to the first band and is negligible  for the second band($m_1 \not= 0, m_2 \approx 0$ and $e \not= 0$). With further increase in $G$ the  magnetization of the first band increases slowly while the magnetization of second band approaches zero, and the strain exhibits near linear growth. These are all due to the redistribution of electrons from the second band to the first band. All these features can be understood from the qusiparticle density of states plotted in Fig.3. The topmost part of DOS which corresponds to $G<G_c$ shows that the DOS is the same for both bands(1 and 2) for a given spin direction. The spectral weights are, however, different for the two spin directions so that we get a non-zero magnetization for both the bands and at the same time, the strain is zero due to equal population of bands 1 and 2.  The position of the  upper Hubbard bands is determined by magnitude of $U$ and since $U$ is large in our case, they are never occupied at  $T=0$ but the spectral weights of the two bands are determined by the sum rule. In this sense the upper Hubbard bands influence the physics of the problem. For example the DOS of spin -up in upper Hubbard band is nearly a $\delta$-function whose weight depends on the electron concentration. For $G > G_c$ the band 1 is more populated than the band 2 leading to a finite strain. The spin-up and spin-down parts of the band 1 are unequally filled leading to a finite magnetization $m_1$. However, for the band 2, the DOS for spin-up and spin-down are the same (dotted line for the spin-down overlaps with full-line for the spin-up case) and this makes $m_2=0$.

\begin{figure}[htb]
\begin{center}
    \epsfig{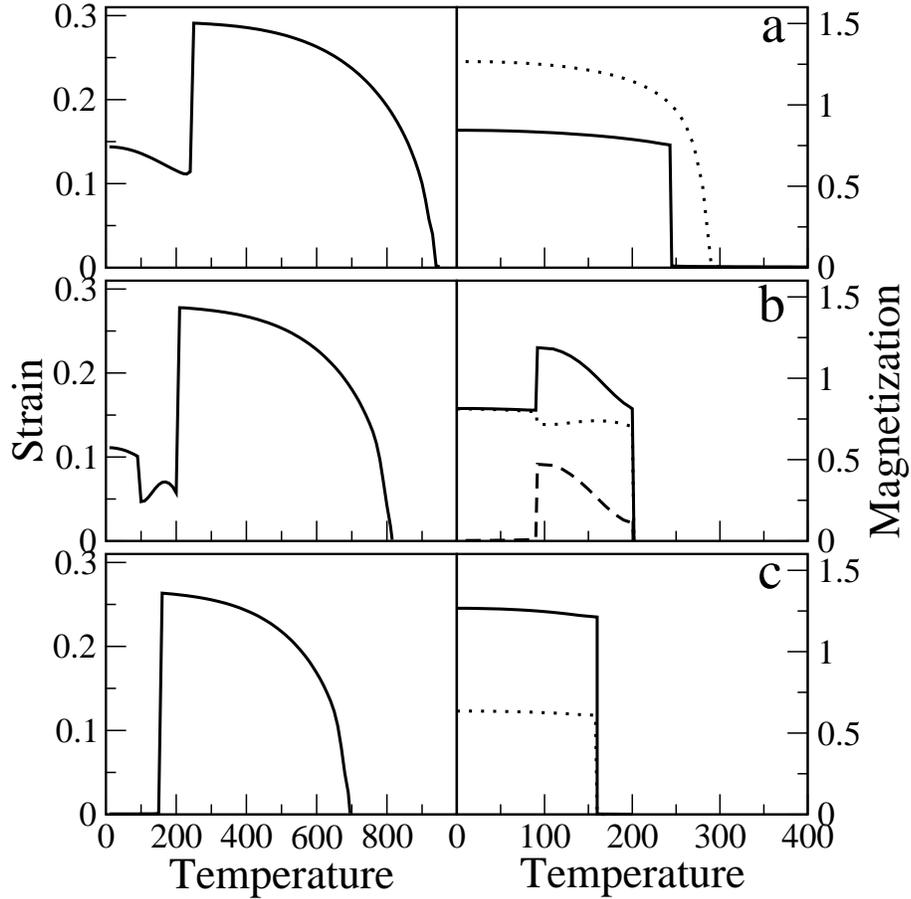}
    \caption{Temperature dependence of strain and magnetization for various values of $G$. $n = 1.4$ and $U = 3$. a) The dotted line is the magnetization in the absence of J-T distortion. $G = 0.49$. b) The dotted and dashed lines are the magnetization of sub bands $1$and $2$ respectively. $G = 0.47$. c) Both the sub bands of equal magnetization which is denoted by the dotted line. $G = 0.45$.}
\end{center}
\end{figure}

\begin{figure}[htb]
\begin{center}
    \epsfig{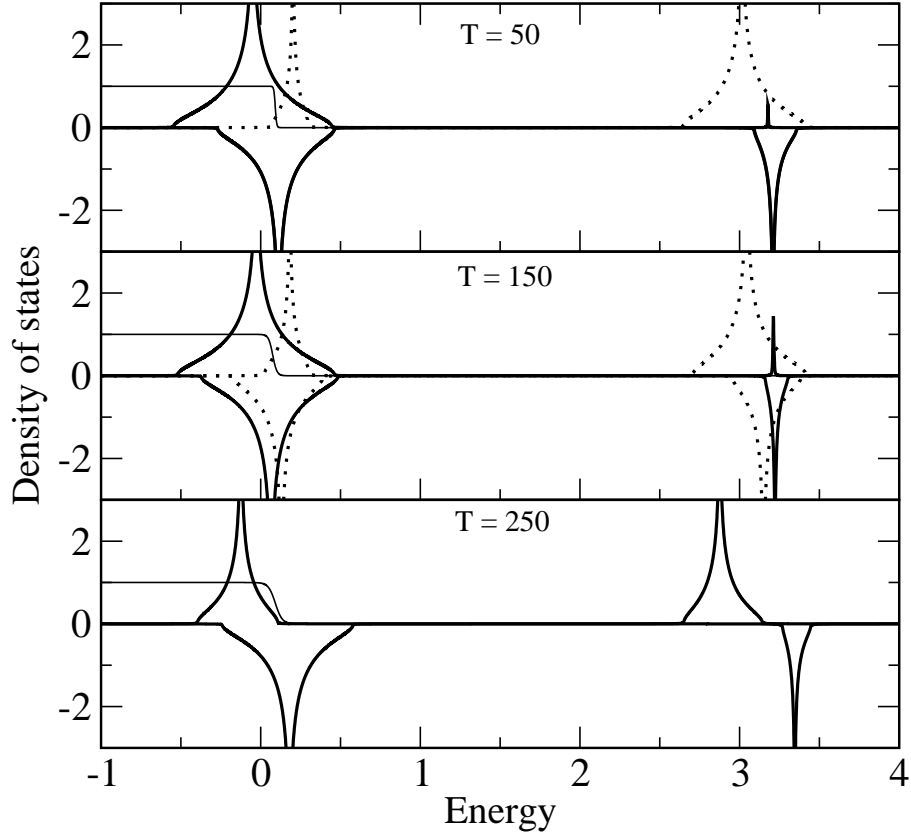}
    \caption{Quasiparticle density of states(for the lower subband in the positive half of the frame and for the upper subband in the negative half of the frame) as a function of energy at various temperatures for $G = 0.47$. Full lines for spin up and dotted lines for spin down. Thin line is the Fermi function. $n = 1.4$, $U = 3$ and $T=0$.}
\end{center}
\end{figure}

From the study of the $G$ dependence of the strain at $T=0$ (Fig.2), it can be concluded that the presence of magnetization causes a redistribution of electrons between the bands by creating a population difference between the spin levels. This redistribution suppresses the strain. In order to study the interplay  between structural and magnetic transitions, we study the temperature dependence of the strain and the spontaneous magnetization for different values of $G$, namely (i) $G > G_c$, (ii) $G > G_c$ but close to $G_c$ and (iii) $G < G_c$ but close to $G_c$ for $U=3$ and $n=1.4$. We note that $G_c=0.462$ for the above parameters. For $G > G_c$, at $T=0$, as pointed out earlier, $m_1 \not= 0, m_2 \approx 0$ and $e \not= 0$. With the increase of temperature, both $m_1$ and $e$ decrease (Fig.4a) slowly at first because of redistribution electrons among spin  and orbital levels of bands. At $T_c$, the magnetization falls to zero abruptly and $e$ rises in a discontinuous manner. It is interesting to note that in the absence of J-T interaction, the magnetization vanishes smoothly at a higher value of $T_c$ (dotted line in right panel of Fig.4a). The J-T interaction suppresses $T_c$ and also the magnitude of magnetization which indicates that the two order parameters compete. Above $T_c$, the strain is high and vanishes smoothly at $T_s$, the transition temperature from distorted to undistorted phases. For $0<T<T_c$ there is coexistence of magnetic order and lattice distortion.  As $T\stackrel{>}{\rightarrow}T_c$ the free energy of the  state with lattice distortion is lowered than that of coexistence phase. We note that the coexistence of magnetic order and lattice distortion has been observed in $La_{1-x}Ca_{x}MnO_{3}$ at $x = 0.15$ [20]. Similar results have been reported [21] recently for the two-band Kondo-lattice model by using an interpolating self-energy approach with a modified Ruderman-Kittel-Kasuya-Yosida mechanism.

When $G > G_c$ but close to $G_c$, the behaviour of $e$ and $m$ are different (Fig.4b) compared to for $G > G_c$ at finite but low temperature. To start with, the magnetization $m_1 \not= 0$, but $m_2 \approx 0$ and $e \not= 0$. However, $e$ is smaller than in the case of $G > G_c$ at low $T$. When the temperature reaches a certain value, $e$ abruptly drops down, $m_1$ decreases slightly and $m_2$ becomes substantially different from zero so that the total magnetization increases sharply. With further increase of $T$,  the magnetization disappears at $T_c$ in a discontinuous fashion and $e$ is also augmented in a similar way. Thus, it is found that a thermal enhancement of magnetization can exist within a certain range of temperature. This interplay of magnetization and distortion is intimately related to the non-linear dependence of spin -level splitting on the J-T-splitting. In other words, the J-T interaction renormalizes the correlation in such a way that the correlation decreases in band 2 which is shifted to higher energy due to the J-T interaction. In such a situation, the magnetization may disappear in this band. However, when the J-T-splitting becomes smaller at higher $T$, the effective correlation in the band 2 goes up leading to a finite magnetization. These arguments are supported by the thermal behaviour of the DOS (Fig.5). At low $T$ ($T=50$) the states  with both the spin directions of band 2 are equally populated and therefore $m_2=0$. On the other hand, in case of the band 1, there is unequal population of the spin sublevels as is evident from the position of the Fermi edge. As $T$ increases, for example at $T=150$, there is a relative shift of the spin-down DOS of the band 2 so that $m_2$ become finite. There is also a small decrease in $m_1$.  Again at $T=250$, the DOS for both the spin directions is the same so that the total magnetization $m=0$ since the spin-down population has increased compared to what it was at $T=50$.

For the case $G < G_c$ but close to $G_c$ (Fig.4c), both the bands  are magnetic and the strain is zero for $0<T<T_c$. Above $T_c$, the strain appears in a discontinuous fashion. When $G$ is decreased further, the strain is always zero for the whole temperature region, even though $G \ne 0$. The re-entrant behaviour of $e$ shown in Fig.4c is observable only in a narrow range of $G$ values close to $G_c$. When $G$ is below this range, the magnetic state is uneffected by J-T interaction.  This demonstrate that the coexistence and mutual competition exist only in a narrow range of parameter space. 

\section{Conclusions}
The J-T interaction is added to the Hubbard model to study the effect of magnetic order on structural distortion in a strongly correlated electron system. The J-T interaction is absorbed into the Hubbard model by renormalizing the band energies. Then the Hubbard model is solved using the spectral density approximation. Using this approximation, the spontaneous strain and magnetization are calculated selfconsistently for various sets of model parameters. The results reported earlier, namely, that ferromagnetic order requires a critical Coulomb repulsion $U$ and is present only for a certain range of band filling [10], and that there is a spontaneous splitting of the degenerate bands only for a J-T coupling $G$ larger than a critical value [21] are recovered. In the strong coupling limit and for small values of $G$ the magnetic state is unaltered in a narrow range of $G$ for a given value of $U$ and $n$ and there is a competition between magnetic order and J-T distortion when they coexist. This is manifested in a reduction of $T_c$ and magnetization.  Within a finite temperature interval an enhancement of magnetization in the magnetic state is found. The general features of these results are in tune with the experimental results in doped manganites [20].

\section{Acknowledgments}

The authors (G.G.R, A.R.K and S.K.G) are grateful to the Council of Scientific and Industrial Research, New Delhi, India for financial support 
through Grant No.03(1068)/06/EMR-II of a scheme.

\section{References}

\noindent [ 1] S. Jin, T.H. Tiefel, M. Mc Cormack, R.A. Fastnacht, R. Ramesh and L.H. Chen, Science \textbf{264} (1994) 413.

\noindent [ 2] A.P. Ramirez, J. Phys.: Condens. Matter \textbf{9} (1997) 8171.

\noindent [ 3] R. Englman, The Jahn-Teller Effect in Molecules and Crystals, Wiely Interscience, New York, 1972.

\noindent [ 4] Y. Tokura, H. KuWahara, Y. Moritomo, Y. Tomioka and A. Asamitsu,
Phys. Rev. Lett. \textbf{76} (1996) 3184.

\noindent [ 5] D.K. Ray and S.K. Ghatak, Phys. Rev. \textbf{B 36} (1987) 3868.

\noindent [ 6] H. Ghosh, M. Mitra, S.N. Behera and S.K. Ghatak, Phys. Rev.
\textbf{B 57} (1998) 13414.

\noindent [ 7]J.D. Fuhr, Michel Avignon and B. Alscio, Phys. Rev. Lett. \textbf{100} (2008) 216402.

\noindent [ 8] J.C. Suits, Solid State Commun. \textbf{18} (1976) 423. 

\noindent [ 9] Shinpei Fhjii, Shoji Ishida and Setsuro Asano, J. Phys. Soc. of Japan \textbf{58} (1989) 3657.

\noindent [10] T. Herrmann and W. Nolting, J. Magn. Mater. \textbf{170} (1997) 253.

\noindent [11] A.B. Harris, R.V. Lange, Phys. Rev. \textbf{157} (1967) 295.

\noindent [12] W. Nolting, M. Potthoff, T. Herrmann and T. Wegner, \textit {in Band-Ferromagnetism Ground-State and Finite-Temperature Phenomena of Lecture Notes in Physics}, Edited by  K. Baberschke, M. Donath and W. Nolting (Springer, Berlin, 2001), p. 208.

\noindent [13] T. Herrmann and W. Nolting, Phys. Rev. \textbf{B 53} (1996) 10579.

\noindent [14] J. Beenen and D.M. Edwards, Phys. Rev. \textbf{B 52} (1995) 13636.

\noindent [15] L.M. Roth ,Phys. Rev. \textbf{184} (1969) 451.

\noindent [16] B. Mehlig, H. Eskes, R. Hayn, and M.B.J. Meinders, Phys. Rev. \textbf{B 52} (1995) 2463.

\noindent [17] H. Mori, Prog. Theor. Phys., \textbf{33} (1965) 423.

\noindent [18] R. Zwanzig, Phys. Rev. \textbf{124} (1961) 983.

\noindent [19] R. Jellito, J. Phys. Chem. Solids \textbf{30} (1969) 609. 

\noindent [20] M. Pissas, I. Margiolaki, G. Papavassiliou, D. Stamopoulos, and D. Argyriou, Phys. Rev. \textbf{B 72} (2005) 064425.

\noindent [21] M. Steir and W. Nolting, Phys. Rev. \textbf{B 75} (2007) 144409.

\end{document}